\documentclass[aps,pra,nofootinbib,superscriptaddress,twocolumn]{revtex4-1}
\usepackage{subfigure,graphicx} 
\usepackage{amsfonts,amssymb,amsmath}
\usepackage{flafter}
\usepackage{multirow}
\usepackage{txfonts}
\usepackage[unicode,breaklinks]{hyperref}
\usepackage{siunitx}
\newcommand{\papertitle}{Simulating cosmological supercooling with a cold atom system}
\hypersetup{
    unicode=true,
    a4paper=true,
    plainpages=false,
    pdftitle={\papertitle},
    pdfauthor={Thomas P. Billam, Kate Brown, Ian G. Moss},
    pdfsubject={\papertitle},
    colorlinks=true,
    linkcolor=blue,
    citecolor=blue,
    filecolor=black,
    urlcolor=blue
}

\usepackage{ulem}

\newcommand{\be}{\begin{equation}}
\newcommand{\ee}{\end{equation}}
\newcommand{\bea}{\begin{eqnarray}}
\newcommand{\eea}{\end{eqnarray}}
\newcommand{\beal}{\begin{aligned}}
  \newcommand{\eeal}{\end{aligned}}

\newcommand{\eqnrefp}[1]{{[Eq.~(\ref{#1})]}}

\newcommand{\figreft}[2]{Fig.~\ref{#1}#2}

\newcommand{\figreftfull}[2]{Figure~\ref{#1}#2}

\begin{document} 

\title{\papertitle}

\author{Thomas P.\ Billam}
\email{thomas.billam@ncl.ac.uk}
\affiliation{Joint Quantum Centre (JQC) Durham--Newcastle, School of Mathematics, Statistics and Physics, 
Newcastle University, Newcastle upon Tyne, NE1 7RU, UK}

\author{Kate Brown}
\email{k.brown@ncl.ac.uk}
\affiliation{School of Mathematics, Statistics and Physics, 
Newcastle University, Newcastle upon Tyne, NE1 7RU, UK}

\author{Ian G. Moss}
\email{ian.moss@ncl.ac.uk}
\affiliation{School of Mathematics, Statistics and Physics, 
Newcastle University, Newcastle upon Tyne, NE1 7RU, UK}

\date{\today}

\begin{abstract}
We perform an analysis
of the supercooled state in an analogue to an early universe phase transition based on a
one dimensional, two-component Bose gas. We demonstrate that the thermal
fluctuations in the relative phase between the components are characteristic of
a relativistic thermal system. Furthermore, we demonstrate the equivalence of
two different approaches to the decay of the metastable state: specifically a
non-perturbative thermal instanton calculation and a stochastic
Gross--Pitaevskii simulation.
\end{abstract}

\maketitle

\section{Introduction}
In its early stages, our universe was filled with hot, relativistic
plasma that cooled through all of the major energy thresholds of fundamental
particle physics, undergoing several changes of phase between different
physical regimes. At the most extreme, the universe may have undergone first 
order transitions, characterised by metastable, supercooled states and the 
nucleation of bubbles. Bubbles of a new matter phase would
produce huge density variations, and unsurprisingly first order phase
transitions have been proposed as sources of gravitational waves
\cite{Caprini:2009fx, Hindmarsh:2013xza} and as sources of primordial black
holes \cite{Hawking:1982ga, Deng:2017uwc}. Despite the importance of
this phenomenon, we have no experimental test of the basic theory.
In this paper, we propose that a thermal supercooled state, analogous to a 
relativistic system, can be realised in a Bose gas experiment.

Phase transitions in fundamental particle physics can be associated with a 
Klein-Gordon field in an effective potential. At high temperatures, the field fluctuates about
the minimum value of the potential representing a high temperature phase. 
As the temperature drops, the minimum of the potential changes to represent the
low temperature phase, but the field can become trapped in a metastable state.
Extreme supercooling can even lead to a zero-temperature metastable  
`false vacuum' state.

The idea of using analogue systems for cosmological processes comes under the
general area of modelling the ``universe in the
laboratory''~\cite{RogerNatCom16,Eckel:2017uqx}. So far, analogue systems have
mostly been employed to test ideas in perturbative quantum field
theory~\cite{Unruh:1980cg, Barcelo:2005fc}, but the non-perturbative phenomenon
of false vacuum decay has recently been discussed, with theoretical descriptions of
vacuum decay of atomic \cite{Braden:2019vsw} and relativistic systems
\cite{Braden:2018tky} at zero temperatures. Among possible analogue systems,
(quasi-)one-dimensional ultracold Bose gases have emerged as an outstandingly
versatile experimental platform for probing many-body quantum dynamics
\cite{LangenExperimental2015, ErneUniversal2018, PruferObservation2018}.

\begin{center}
\begin{figure}[tbh]
  \includegraphics[width=\columnwidth]{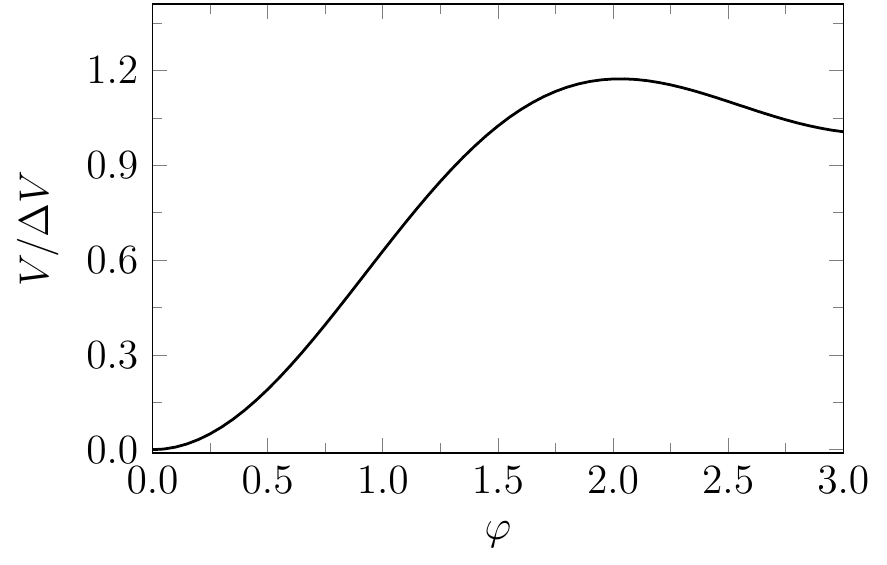}
\caption{The field potential $V$ plotted as a function of the
relative phase of the two atomic wave functions, $\varphi$. The
metastable phase is at the minimum $\varphi=\pi$ and the stable
phase is at the global minimum $\varphi=0$. The difference in energy
density between these phases is $\Delta V$.} \label{pot} 
\end{figure} 
\end{center}

Fialko et al. \cite{FialkoFate2015,FialkoUniverse2017}
proposed an actual experiment to simulate the relativistic vacuum decay in a cold atom system. 
Their system consists of a Bose gas with two different spin states of the same atom species in an
optical trap. The two states are coupled by a microwave field.  By modulating
the amplitude of the microwave field, a new quartic interaction between the two
states is induced in the time-averaged theory which creates a non-trivial
ground state structure as illustrated in \figreft{pot}{}.

In this paper we introduce thermal effects into the model of Fialko et al. to better
replicate the conditions relevant to the very early universe. We take their time-averaged
potential and study the physics of supercooling for a metastable state
in the thermal ``cross-over'' regime of a quasi-one-dimensional Bose gas.
We demonstrate the bubble nucleation dynamics of the first order
phase transition are correctly reproduced by numerical modelling using a stochastic projected
Gross--Pitaevskii equation (SPGPE)~\cite{GardinerStochastic2002,
GardinerStochastic2003, bradley_bose-einstein_2008, Blakie2008,
BradleyStochastic2014}.
We show that the stochastic approach
agrees with semi-classical predictions based on non-equilibrium thermal field
theory of a relativistic Klein Gordon system. This agreement applies not only to the correlation functions,
but also to the non-perturbative decay rate of a metastable state.

This paper also sets up further work extending the results to oscillating potentials which we
hope to present in due course. Braden et al. have shown that oscillating potentials
lead to a parametric instability on small wavelengths which causes a classical decay 
of the metastable state \cite{Braden:2017add,Braden:2018tky}. In order to produce a 
supercooled system, some form of dissipative mechanism would have to act on small scales 
to damp out the parametric resonance, and we propose that thermal damping may
be the solution. 

\section{System}
Our system is a one-dimensional, two-component Bose gas of atoms with mass $m$.
The two components are different spin states of the same species, coupled by a
time-modulated microwave field. The Hamiltonian is given by
\begin{equation}
H=\int dx\left\{
-{\hbar^2\over 2m}\psi^\dagger\nabla^2\psi
+V(\psi,\psi^\dagger)\right\},
\label{hamil}
\end{equation}
where the field operator  $\psi$ has two components $\psi_i$, $i=1,2$.
Fialko et al. \cite{FialkoFate2015,FialkoUniverse2017} 
described a procedure whereby averaging 
over timescales longer than the modulation timescale
can lead to an interaction potential of the form
\begin{equation}
V=\frac{g}2\sum_i\left(\psi_i^\dagger\psi_i\right)^2 - \mu \psi^\dagger \psi
-\mu\epsilon^2\psi^\dagger\sigma_x \psi+
\frac{g}{4}\epsilon^2\lambda^2(\psi^\dagger\sigma_y\psi)^2,\label{potential}
\end{equation}
where $\sigma_{\{x,y\}}$ are Pauli matrices.
The potential includes the chemical potential $\mu$, 
equal intra-component $s$-wave interactions of strength $g$ between the field operators 
(we assume inter-component $s$-wave interactions are negligible), and a microwave-induced 
interaction with strength $\mu\epsilon^2$. The final term comes from the 
averaging procedure, and introduces a new parameter $\lambda$, dependent
on the amplitude of the modulation.
The trapping potential used to confine the condensate has been omitted in 
order to isolate the physics of vacuum decay. In principle, a quasi-one-dimensional
ring trap experiment could realize the uniform system we study.

The terms proportional to $\epsilon^2$ are responsible for the difference in energy 
between the global and local minima of the energy, and we require
$\epsilon$ to be small. The global minimum represents the true vacuum state 
and the local minimum represents the false vacuum.
The true vacuum is a state with $\psi_1=\psi_2$ and the
false vacuum is a state with $\psi_1=-\psi_2$. The condensate densities 
of the two components at the extrema are equal to one another, and 
given by $\langle \psi_1^\dagger\psi_1\rangle =\langle\psi_2^\dagger\psi_2\rangle 
=\rho_0(1\pm\epsilon^2)$,  in terms of the mean 
density $\rho_0=\mu/g$.

Throughout this paper, we will make use of the healing length
$\xi=\hbar/(mg\rho_0)^{1/2}$ and the sound speed $c=\hbar/(m\xi)$. 
Together, these define a characteristic frequency $\omega_0=c/\xi$.
The dimensionless form of the potential constructed from these parameters becomes
$\hat V=V/(\hbar\omega_0\rho_0)$.
If we now introduce the relative phase $\varphi$ between the spin components,
such that $\psi_1\approx\rho_0 e^{i\varphi/2}$
and $\psi_2\approx\rho_0 e^{-i\varphi/2}$, then the potential becomes
\begin{equation}
\hat V\approx-2\epsilon^2-2\epsilon^2\cos\varphi+\epsilon^2\lambda^2\sin^2\varphi,
\label{vhat}
\end{equation}
as shown in \figreft{pot}{}, with $\Delta V=4\hbar\omega_0\rho_0\epsilon^2$. 

In the experimental proposal, the system is initially prepared in the metastable
phase at a temperature $T$. In one dimension, the physics of Bose gases
critically depends on the dimensionless interaction strength parameter $\zeta =
(\rho_0 \xi)^{-2}$ and the temperature \cite{KheruntsyanPair2003,
  KheruntsyanFinite2005, BouchouleTwoBody2012, HenkelCrossover2017}. We consider
the weakly interacting case, $\zeta \ll 1$. A phase-fluctuating
quasi-condensate, in which density fluctuations are suppressed, appears at
temperatures below the cross-over temperature \footnote{Note that our definition
  omits a numerical factor 2 often found elsewhere~\cite{KheruntsyanPair2003,
    KheruntsyanFinite2005, BouchouleTwoBody2012}.}
\begin{equation}
T_{CO}={\hbar c\rho_0\over k_B}.
\end{equation}
The gas remains degenerate up to a temperature of order
$T_D=\zeta^{-1/2}T_{CO}>T_{CO}$.

\section{Stochastic Gross Pitaevskii equation}
Stochastic Gross--Pitaevskii equations (SGPEs) are widely used for modeling
atomic gases at and below the condensation temperature~\cite{StoofCoherent1999,
StoofDynamics2001, GardinerStochastic2002, GardinerStochastic2003,
bradley_bose-einstein_2008, Blakie2008}. Here, we use the \textit{simple growth}
stochastic projected Gross--Pitaevskii equation
(SPGPE)~\cite{bradley_bose-einstein_2008, Blakie2008}, which has been
successfully used to model experimental phase
transitions~\cite{weiler_spontaneous_2008, liu_dynamical_2018}. Extension of the
SPGPE to spinor and multi-component condensates is described in
Ref.~\cite{BradleyStochastic2014}.

For convenience, from this point in the paper we use $\xi$ as the length unit
and $\omega_0^{-1}$ as the time unit. We also rescale the wave function by
replacing $\psi\to\rho_0^{1/2}\psi$, and measure the temperature in units of
$T_{CO}$. In these units, the form of SPGPE we use is
\begin{equation}
i{\partial\psi_j\over\partial t}={\cal P}\left\{(1-i\gamma)\left(-\frac12\nabla^2\psi_j+
{\partial \hat V\over\partial  \psi_j^\dagger}\right)+\eta_j\right\}.
\label{sgpe}
\end{equation}
Here the complex fields $\psi_j$ describe the well-occupied, low-momentum modes
of the system (the c-field region), and the projector $\mathcal{P}$ eliminates
modes above the momentum cut-off $k_{\mathrm{cut}} = \sqrt{2 \rho_0 \xi T}$. We
also considered other values of $k_{\textrm{cut}}$, to ensure our results were
not overly sensitive to the choice of momentum cut-off. The noise source
$\eta$ is a Gaussian random field with correlation function
\begin{equation}
\langle\eta_i(x,t)\eta_j(x',t')\rangle=2\gamma T\delta(x-x')\delta(t-t')\delta_{ij},
\end{equation}
and the potential
\begin{equation}
\hat V=\frac12\sum_i\left(\psi_i^\dagger\psi_i-1\right)^2-\epsilon^2\psi^\dagger\sigma_x\psi+
\frac14\lambda^2\epsilon^2(\psi^\dagger\sigma_y\psi)^2 .
\end{equation}
Typically, we set the dimensionless
dissipation rate $\gamma = 10^{-2}$. Values of $\gamma$ from
$\mathcal{O}(10^{-4})$ to $\mathcal{O} (10^{-2})$ have been used in previous
work that made direct comparisons to experiment~\cite{weiler_spontaneous_2008,
rooney_persistent-current_2013, ota_collisionless_2018, liu_dynamical_2018},
making this a reasonable choice. We comment on the effect of $\gamma$ later in
the text. Our SPGPE simulations use a one dimensional grid of size $L=240\xi$ with
periodic boundaries and spacing $\Delta x = 0.4\xi$. We set $\rho_0 \xi = 100$.
Our simulations were executed using the software package XMDS2
\cite{DennisXMDS2013}. Averaged quantities were calculated over $1000$
stochastic realizations.

Some care should be exercised when applying the SPGPE in reduced dimensions,
since a three-dimensional thermal cloud is
assumed~\cite{BradleyLowDimensional2015}. For a gas confined in a transverse
harmonic trap of frequency $\omega_\perp$, the simple growth SPGPE above is
valid in 1D with dimensionally-reduced interaction strength $g = 2\hbar a_s
\omega_\perp$ provided $\hbar \omega_\perp \lesssim k_\mathrm{B} T$ and, in
principle, $\mu \ll \hbar \omega_\perp$. In practice $\mu \lesssim \hbar
\omega_\perp$ is sufficient: 1D S(P)GPE equilibrium states were investigated in
Refs.~\cite{CockburnQuantitative2011, DavisYangYang2012} and shown to be an
excellent match to quasi-1D atom-chip experiments in this regime
\cite{TrebbiaExperimental2006, vanAmerongenYangYang2008}. 

We will be running at some fraction of the cross-over temperature
$T_{CO}$ which is larger than the temperature, $T_\phi = \hbar^2 \rho_0 / (m k_{\mathrm{B}} L) $, 
at which phase coherence is attained across the entire system
Crucially however, the relative phase $\varphi$ has an effective potential barrier that 
assists phase coherence in the relative phase at higher temperatures than $T_{\phi}$, as we
shall see in the following results.

\begin{center}
  \begin{figure}[htb]
    \includegraphics[width=\columnwidth]{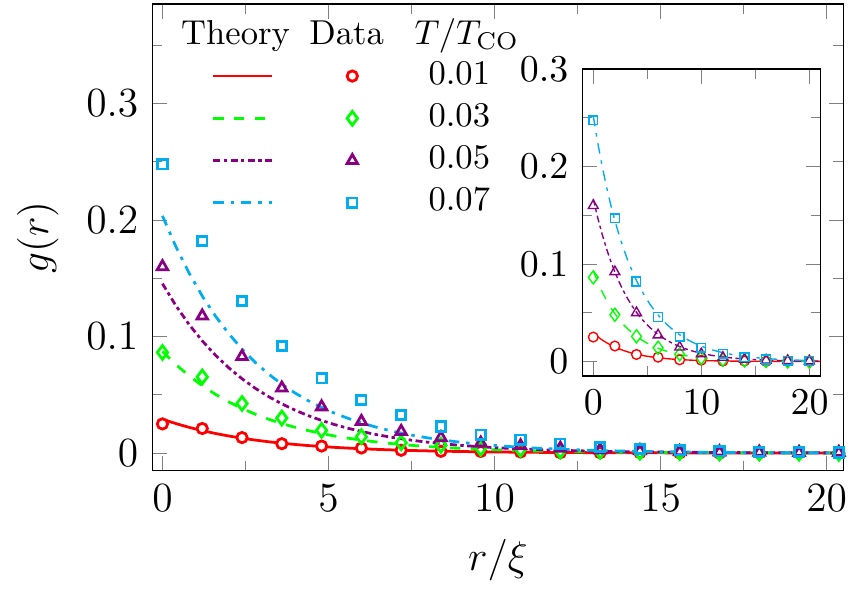}
    \caption{The correlation function for the relative phase, $g(|x-x'|)=\langle \varphi(x,t)\varphi(x',t)\rangle$,
      at different temperatures in a potential with $\lambda=1.4$ and $\epsilon=0.1$.
      The SPGPE results agree with the correlations of a thermal Klein Gordon field at
      the lower temperatures, but non-linear effects start to cause a difference as the 
      temperature increases. Nonetheless, better agreement can be achieved at
      higher temperatures by fitting an \textit{effective} value of 
      $\lambda_{\mathrm{eff}}$ at each temperature (inset).
    } \label{cor} 
  \end{figure} 
\end{center} 


\section{Results}
\subsection{Equilibrium correlations}
The correlation functions for fluctuations about the stable and metastable
phases provide an essential check on the validity of the numerical modelling,
and also elucidates the relation between fluctuations in the SGPE and the
Klein-Gordon field $\varphi$. As shown in Appendix~\ref{appendix:correlations},
small fluctuations in the relative phase $\varphi$
of the two components induced by the SGPE have a thermal Klein Gordon
correlation function
\begin{equation} \langle \varphi(x,t)\varphi(x',t)\rangle= {T\over
m_\varphi}e^{-m_\varphi |x-x'|} + {T\over 2}\delta(x-x'),
\end{equation} where the Klein-Gordon mass $m_\varphi=2\epsilon(\lambda^2\pm
1)^{1/2}$, for the stable and metastable phases respectively.

The correlation function about the stable phase computed from SPGPE simulations
is shown in \figreft{cor}{}. As expected, at low temperatures we have complete
agreement with the Klein-Gordon result. At higher temperatures, non-linear
effects are increasingly important, until the state becomes completely phase
incoherent, in analogy to symmetry restoration in fundamental particle physics.
At intermediate temperatures, we can restore the agreement against the
theoretical result by introducing an `effective' coupling' $\lambda_{\rm eff}$,
as shown in the inset in \figreft{cor}{}.

\subsection{Bubble nucleation}
In a first order regime, we expect to see exponential decay of the metastable
state, triggered by bubble nucleation events. In this section we present
numerical results which confirm this prediction, and we show agreement with a
semi-classical, non-perturbative instanton approach.

\begin{center}
\begin{figure}[htb]
  \includegraphics[width=\columnwidth]{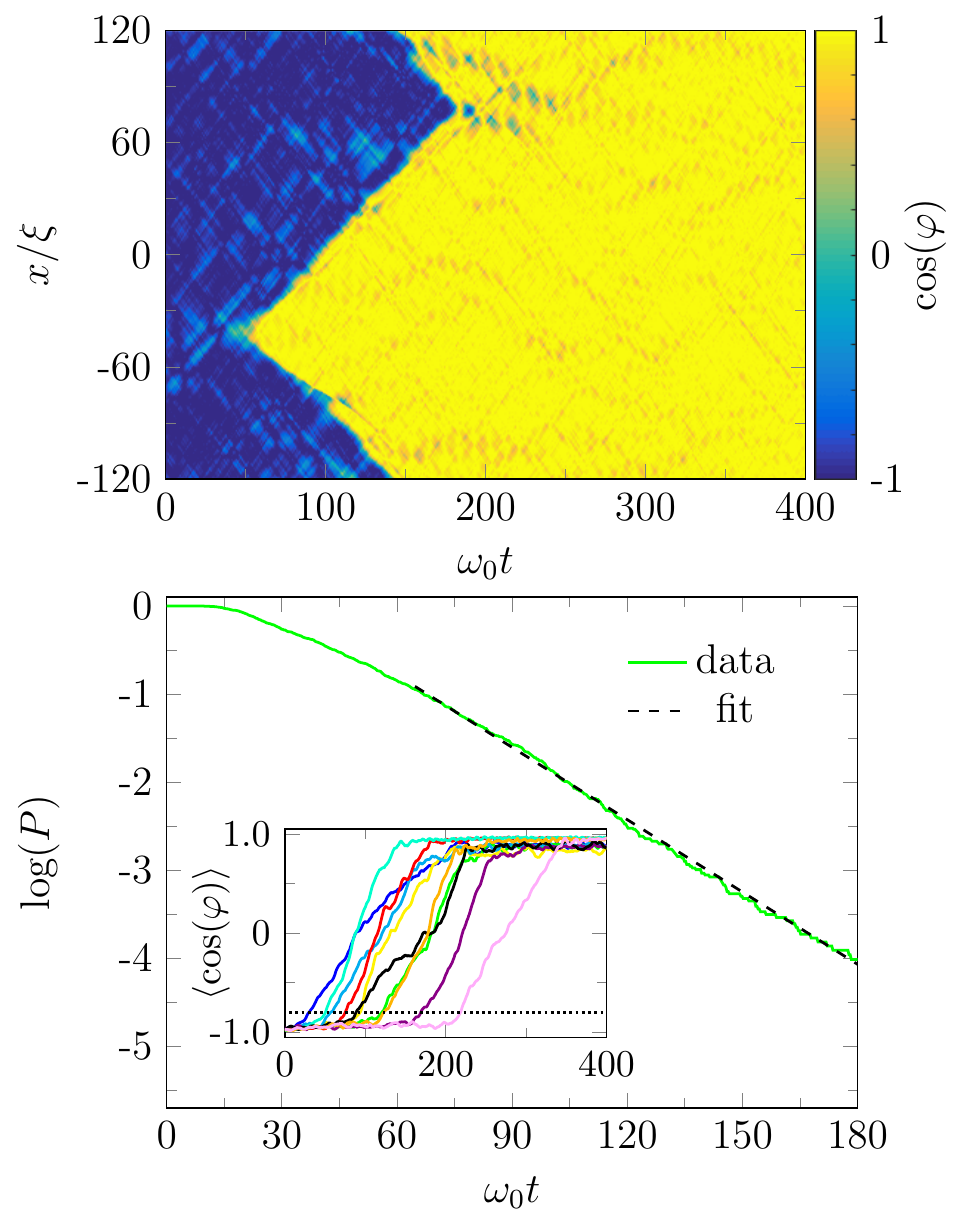}
\caption{Above, an example of bubble nucleation for $\lambda=1.4$ and $T=0.03T_{CO}$. Below,
the logarithm of the probability, $P$, of remaining in the metastable state.
In this case the system was first allowed to equilibrate at $\lambda=1.8$,
a barrier large enough that bubble nucleation was negligible, after
which the barrier was reduced to $\lambda=1.4$. Inset, the spatial average of $\cos\varphi$ 
for ten different runs. The nucleation time is taken to be when 
$\langle \cos \varphi \rangle > -1+\Delta$, 
where $\Delta=0.2$ in this example.} 
\label{bubbles} 
\end{figure} 
\end{center}

\begin{center}
  \begin{figure}[htb]
    \includegraphics[width=\columnwidth]{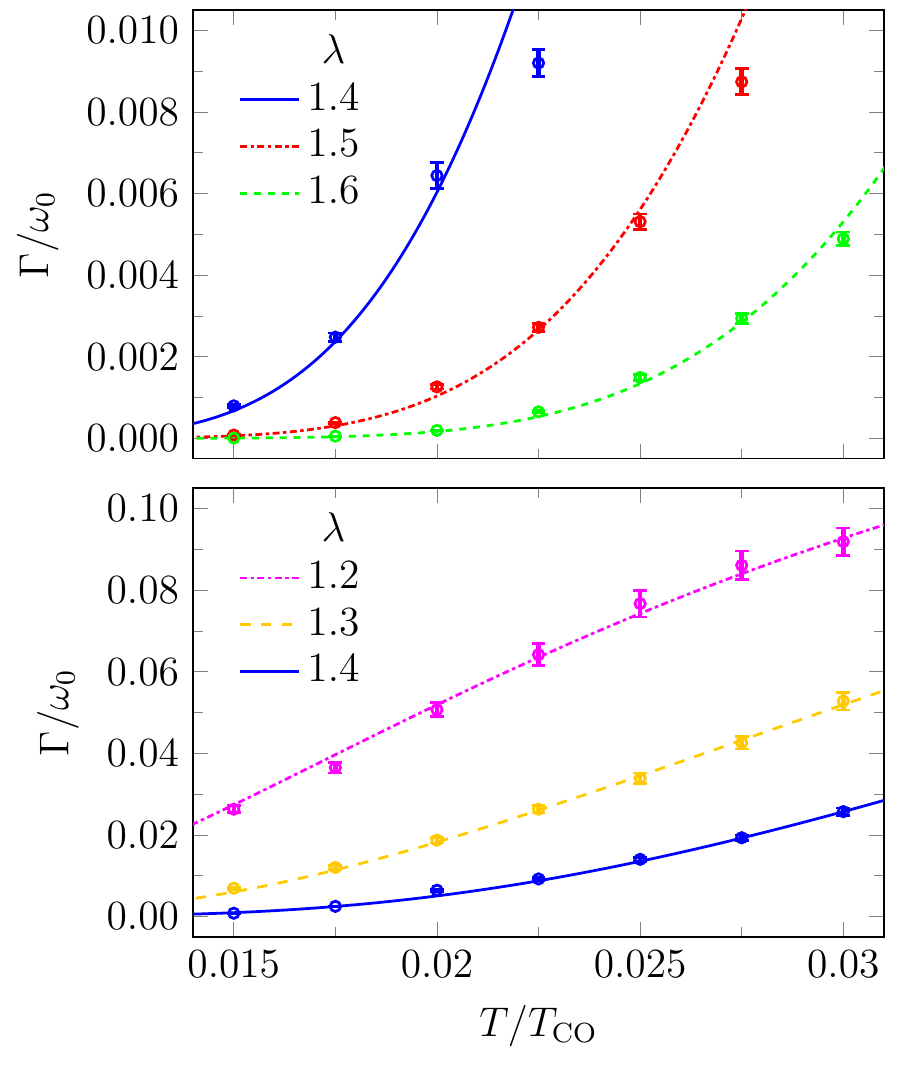}
    \caption{A comparison between the decay constant obtained from the SPGPE
      (data points) and the
      instanton method (lines) as a function of temperature. These plots are for $\epsilon=0.1$, interaction
      strength $\zeta=10^{-4}$ and dissipation $\gamma=10^{-2}$. There is good agreement when
      the decay constant  $\Gamma<\gamma$, and this can be extended to higher $\Gamma$ by
      using an effective coupling $\lambda_{\rm eff}$ (lower plot).
    } \label{compare} 
  \end{figure} 
\end{center}

In order to model bubble nucleation using the the SPGPE, we must initialize the system in the
metastable state. In most runs we do this by placing the fields $\psi_j$
in the fluctuation-free metastable state at time $t=0$, allowing the noise term
in the SPGPE \eqnrefp{sgpe} to rapidly build up thermal fluctuations.
We also verified that equivalent results are produced by first allowing the fields to thermalize
with a high potential barrier ($\lambda=1.8$) and then instantaneously
reducing $\lambda$, since this latter procedure is 
closer to a likely experimental protocol. A signature of bubble formation
in an individual trajectory is given by the spatial average
$\langle\cos\varphi\rangle$ exceeding $-1+\Delta$, where $\Delta=0.2$ is chosen to
be larger than the typical fluctuations of $\langle\cos\varphi\rangle$ due to
thermal noise in the system. An example is shown in \figreft{bubbles}{}. Running
many stochastic trajectories and computing the probability, $P$, of remaining in
the metastable state results in an exponential decay curve, also shown in
\figreft{bubbles}{}. A fit to the exponential form $P=a e^{-\Gamma t}$ over the
time intervals seen to be exhibiting exponential decay yields the
decay rate $\Gamma$. \figreftfull{compare}{} shows the decay rate $\Gamma$ for several values of $T$
and $\lambda$. Uncertainties on $\Gamma$, reflecting the statistical
uncertainty arising from the trajectory averaging, are computed by a bootstrap
resampling approach~\cite{BillamSimulating2019}.

The semi-classical model of bubble decay is based on an instanton calculation, where the equations
are solved in imaginary time $\tau$ to give an instanton solution $\psi_b$. For thermal scenarios, 
the imaginary time coordinate is taken to be periodic, with period $\beta=\hbar/(k_BT)$.
The instanton solution approaches the metastable state at large distances, and for
a purely thermal transition the solution is independent of $\tau$.

The full expression for the 
nucleation rate of vacuum bubbles in a volume ${\cal V}$ is
\cite{Coleman:1977py,Callan:1977pt},
\begin{equation}
\Gamma \approx {\cal V}A_0B^{1/2}e^{-B}.\label{gamma}
\end{equation}
where $B$ denotes the difference in action between the instanton and the
metastable state divided by $\hbar$. The pre-factor $A_0$ depends on the change in the spectra
of the perturbative modes induced by the instanton. This should only depend
mildly on temperature, so we will treat this term as an undetermined constant.

The exponent is explicitly
\begin{equation}
B=\rho_0\int_{-\infty}^\infty dx \int_0^\beta d\tau\left\{\psi_b^\dagger{\partial \psi_b\over\partial\tau}
+\frac12\psi_b^\dagger\nabla^2\psi_b+\hat V\right\}.
\end{equation}
In the Klein-Gordon approximation, the decay exponent simplifies to
\begin{equation}
B={\alpha(\lambda)\epsilon\over T},
\end{equation}
where the factor $\alpha(\lambda)$ is defined by
\begin{equation}
\alpha(\lambda)=
{1\over 4\epsilon}\int_{-\infty}^\infty dx\left\{\left({\partial\varphi_b\over\partial x}\right)^2+4\hat V\right\}.
\label{defalpha}
\end{equation}
Note that the $\epsilon$ dependence in (\ref{defalpha}) disappears if we rescale $x\to x/\epsilon$ and use Eq. (\ref{vhat}). 

The values of $\alpha(\lambda)$ for a Klein-Gordon model have been obtained recently in Ref. \cite{Mario}.
A comparison between the instanton and stochastic approaches is shown in \figreft{compare}{}.
They agree very well in the region were $\Gamma<\gamma$, which we interpret as the nucleation rate
having to be less that the relaxation rate of the thermal ensemble. Remarkably, the two approaches also
agree over a wider range if we replace the coupling $\lambda$ by an `effective' 
value $\lambda_{\rm eff}$.

Finally, we note that we repeated a sample of our SPGPE
simulations with lower dissipation rate
$\gamma = 5\times 10^{-3}$. We find that the rate $\Gamma$  is dependent on $\gamma$.
However, the results are still well-fitted by the instanton approach,
but with a different pre-factor $A_0$, as would be expected from the theory
of dissipative tunnelling in quantum mechanics \cite{PhysRevLett.46.211}.

\section{Conclusion}
The quasi-condensed thermal Bose gas described above would serve as a laboratory
analogue to an early universe, supercooled phase transition. We show that the
SPGPE can be used to model the system, and that where overlap with instanton
calculations is possible there is agreement between the predictions of the two
approaches.

As an example experimental configuration, we consider one of the experimental
setups proposed by Fialko et al. \cite{FialkoUniverse2017}, which is based on
tuning the interactions between two Zeeman states of ${}^7{\rm Li}$ \footnote{In
this example the intra-component scattering lengths of the spin states are
asymmetrical and the potential (\ref{potential}) has to be slightly modified
\cite{FialkoUniverse2017}.}. The interactions can be tuned using a Feshbach
resonance to achieve the required close-to-zero inter-component scattering
length~\cite{FialkoUniverse2017}. Based on the average intra-component scattering
length, suitable experimental parameters would be $5 \times 10^4$ atoms in a
quasi-1D optical trap \cite{SalcesCarcobaEquation2018} of length $\SI{90}{\micro
m}$ and transverse frequency $2\pi \times \SI{66}{\kilo Hz}$. The interaction
strength $\zeta=10^{-4}$ (as in \figreft{compare}{}), and the cross-over
temperature $T_{CO}= \SI{215}{\micro K}$. In this context the results in
\figreft{compare}{} correspond to temperatures from around $\SI{3.2}{\micro K}$
to $\SI{6.4}{\micro K}$, where bubble nucleation should be observable.

Interestingly, the phase correlation length at the temperatures of interest
is less than the length of the gas, but the {\it relative} phase correlation length is 
larger than the system. This is because the phase of an individual atom
is not constrained by the interactions; in the language of particle physics
it is a `Goldstone mode'. On the other hand, the relative phase develops a 
potential and behaves as a massive Klein-Gordon field with correlation length
fixed by the mass, as shown in Figure \ref{cor}.

In future work, we will investigate a modulated (i.e., time varying) potential
in order to investigate the effects of thermal damping on parametric instabilities.
The semi-classical model of bubble nucleation used here is no longer applicable
for a modulated potential, but the SPGPE will apply provided that $k_BT$ is 
larger than the maximum energy per mode $\hbar\omega_k$. Parametric resonance
sets in on wavelengths a little less than the correlation length $\xi$
\cite{Braden:2017add,Braden:2018tky}, and $k_BT>\hbar\omega_k$
in the parameter ranges discussed in the present paper. 
The parametric resonance has a fixed growth rate and it will be
damped out given sufficiently large thermal damping.
It is important to determine the exact degree of thermal damping required
to see if this is physically realistic.

Also in future work, we will extend our results to two dimensions and include 
realistic trapping potentials. There is a possibility that the boundaries of the trap 
affect bubble nucleation, as we found in \cite{BillamSimulating2019}, and this 
requires further theoretical investigation before two-dimensional simulations
of vacuum  decay and supercooling are constructed.


Data supporting this publication is openly available under a
Creative Commons CC-BY-4.0 License on the data.ncl.ac.uk
site \cite{data}.

\section*{Acknowledgements}
This work was supported in part by the Leverhulme Trust [grant RPG-2016-233],
the UK EPSRC [grant EP/R021074/1], and STFC [grant ST/T000708/1]. KB is supported
by an STFC studentship.
This research made use of the Rocket High Performance Computing service at 
Newcastle University. 

\appendix

\section{Klein-Gordon reduction of the SGPE}
\label{appendix:correlations}
Reduction to a Klein-Gordon system starts from
\begin{align}
\psi_1&=e^{\chi/2}e^{\sigma/2}e^{i\varphi/2}e^{i\theta/2},\\
\psi_2&=e^{\chi/2}e^{-\sigma/2}e^{-i\varphi/2}e^{i\theta/2},
\end{align}
in an approximation where $\chi$, $\sigma$, $\nabla$ and $\partial_t$ are all $O(\epsilon)$.
When these are inserted into Eq. (\textcolor{red}{5}) of the main text, the system reduces at leading order in $\epsilon$ to
\begin{align}
\dot\varphi&=-2\sigma+\eta_\varphi,\\
\dot\sigma&=-\frac12\nabla^2\varphi-2\gamma\sigma+\partial_\varphi \hat V+\eta_\sigma,\\
\dot\theta&=-2\chi+\eta_\theta,\\
\dot\chi&=-\frac12\nabla^2\theta-2\gamma\chi+\eta_\chi,
\end{align}
where the noise terms have a gaussian distribution with covariance $2\gamma T$.

For small $\varphi$, these are a set of stochastic equations for the Bogliubov modes.
In this limit, with $\hat V=m_\varphi^2\varphi^2/4$, they can be solved using green function techniques.
The linearised equations for the transforms $\hat\varphi(k,\omega)$ and $\hat\sigma(k,\omega)$ are
\begin{equation}
\begin{pmatrix}
-i\omega&2\\
-\omega_k^2/2&-i\omega+2\gamma\\
\end{pmatrix}
\begin{pmatrix}
\hat\varphi \\ \hat\sigma\\
\end{pmatrix}=
\begin{pmatrix}
\hat\eta_\varphi \\\hat\eta_\sigma
\end{pmatrix},
\end{equation}
where $\omega_k^2=k^2+m_\varphi^2$. The inverse of the operator matrix
is the Green function $G$,
\begin{equation}
G=
{1\over\Delta}\begin{pmatrix}
i\omega-2\gamma&2\\
-\omega_k^2/2&i\omega\\
\end{pmatrix}
\begin{pmatrix}
\hat\varphi\\\hat\sigma\\
\end{pmatrix},
\end{equation}
where $\Delta=\omega^2+2i\gamma\omega-\omega_k^2$. The stochastic correllator
of the relative phase fluctuations is related to the Green function by
\begin{equation}
\langle\hat\varphi(k,\omega)\hat\varphi(k',\omega')\rangle=
\delta_{\omega\omega'}\delta_{kk'}2\gamma T\left(
G_{\varphi\varphi}G_{\varphi\varphi}^*+G_{\varphi\sigma}G_{\varphi\sigma}^*
\right)
\end{equation}
Inverting the Fourier transform in $t$ gives the phase space correllator
\begin{equation}
\langle \hat\varphi(k,t)\hat\varphi^*(k',t)\rangle={4+\omega_k^2\over 2\omega_k^2}T\delta_{kk'}.
\end{equation}
The first term in the denominator reproduces the thermal correllator for a free Klein Gordon field,
which has energy of $T$ per Fourrier mode.
In position space,
\begin{equation}
\langle \varphi(x,t)\varphi(x',t)\rangle=
{T\over m_\varphi}e^{-m_\varphi r} + {T\over 2}\delta(r),
\end{equation}
where $r=|x-x'|$. The mass is given by $m_\varphi=2\epsilon(\lambda^2\pm 1)^{1/2}$ for the stable 
and metastable states respectively. 

The result is valid for linearised theory. Including higher order terms makes a difference
at higher temperature.
The next order in perturbation theory introduces $T^2$ terms, or using re-summation modifies 
the mass $m_\varphi^2\to m_\varphi^2+\kappa T$, where $\kappa$ depends on regularisation.

\bibliography{paper}

\end{document}